\documentclass[
reprint,
showpacs,
nofootinbib,
 amsmath,amssymb,
 aps,
pra,
floatfix,
]{revtex4-1}

\usepackage{graphicx}% Include figure files
\usepackage{dcolumn}% Align table columns on decimal point
\usepackage{bm}% bold math
\usepackage[colorlinks=true, citecolor=blue, urlcolor=blue ]{hyperref}% add hypertext capabilities
\usepackage{pxfonts,txfonts}
\newcommand{\bq}{\begin{equation}}
\newcommand{\eq}{\end{equation}}
\newcommand{\bqs}{\begin{equation*}}
\newcommand{\eqs}{\end{equation*}}
\newcommand{\ba}{\begin{array}}
\newcommand{\ea}{\end{array}}
\newcommand{\bas}{\begin{array*}}
\newcommand{\eas}{\end{array*}}
\newcommand{\bqa}{\begin{eqnarray}}
\newcommand{\eqa}{\end{eqnarray}}
\newcommand{\bqas}{\begin{eqnarray*}}
\newcommand{\eqas}{\end{eqnarray*}}

\DeclareMathOperator{\tr}{Tr}

\newcommand{\etal}{\textit{et al.}}

\newcommand{\ran}{\rangle}
\newcommand{\lan}{\langle}

\newcommand{\D}{\mathcal{D}}
\newcommand{\N}{\mathcal{N}}

\newtheorem{theorem}{Theorem}

\newcommand{\qed}{\nobreak \ifvmode \relax \else
\ifdim\lastskip<1.5em \hskip-\lastskip
\hskip1.5em plus0em minus0.5em \fi \nobreak
\vrule height0.3em width0.5em depth0.25em\fi}

\begin{document}

%\preprint{APS/123-QED}

\title{Entanglement is not a lower bound for geometric discord}
\author{Swapan Rana}
\email{swapanqic@gmail.com}
\author{Preeti Parashar}
\email{parashar@isical.ac.in}
\affiliation{Physics and Applied Mathematics Unit, Indian Statistical Institute, 203 B T Road, Kolkata, India}
\date{\today}

\begin{abstract}  We show that partial transposition of any $2\otimes n$ state can have at most $(n-1)$ number of negative eigenvalues. This extends a decade old result of $2\otimes 2$ case by Sanpera \etal  [\href{http://dx.doi.org/10.1103/PhysRevA.58.826}{Phys. Rev. A {\bf 58}, 826 (1998)}]. We then apply this result to critically assess an important conjecture recently made in [\href{http://dx.doi.org/10.1103/PhysRevA.84.052110}{Phys. Rev. A {\bf 84}, 052110 (2011)}], namely, the (normalized) geometric discord should always be lower bounded by squared negativity. This conjecture has strengthened the common belief that measures of \textit{generic} quantum correlations should be more than those of entanglement. Our analysis shows that unfortunately this is not the case and we give several counterexamples to this conjecture. All the examples considered here are in finite dimensions. Surprisingly, there are counterexamples in $2\otimes n$ for any $n>2$. Coincidentally, it appears that the $4\otimes 4$ Werner state, when seen as a $2\otimes 8$ dimensional state, also violates the conjecture. This result contributes significantly to the negative side of the current ongoing debates on the defining notion of geometric discord as a  good measure of generic correlations.
\end{abstract}


\maketitle

%\tableofcontents

\section{\label{intro}Introduction}

Entanglement has been a well known signature of non-classicality and considered to be synonymous with quantum correlations. However, investigations in the recent past have revealed that there is more to quantum correlations than just entanglement. Towards this end, quantum discord was proposed \cite{OllivierandZurekPRL01} to quantify the \textit{quantumness} of a quantum state, finding its usefulness as a resource for efficient simulation of some classically intractable computational tasks \cite{DattaetalPRL08}.  A variant of this, namely the geometric discord (GD), introduced in \cite{DakicPRL10}, has attracted a lot of attention, mainly due to its calculational simplicity. Besides, it has been revealed very recently that GD has several interesting operational interpretations such as its optimality as  a resource for remote state preparation \cite{DakicetalAR12,TufarellietalAr12,ChavesandMeloPRA11}, in quantum random access code protocol \cite{YaoetalAR12} and in terms of capability of modifying a global state minimally by non-trivial local unitary dynamics \cite{StreltsovetalAR12}. However, despite such novel features, the defining notion of GD as a good measure of quantumness has itself been questioned recently  \cite{PianiAR12}. 

The correlations captured by quantum discord are more generic in the sense that they can exist even in the absence of entanglement and in fact have been shown to be non-zero for almost all bipartite states \cite{FerraroetalPRA10}.  Intuitively this indicates that its measure (quantifier) can never be less than any entanglement measure. Indeed it has been shown to hold for a wide class of measures including relative entropy of quantumness, the quantum deficit, and the negativity of quantumness \cite{PianiAdessoPRA12}. This view was all the more strengthened by a recent conjecture made in \cite{GirolamiAdessoPRA11}, where the authors have suggested that (normalized) geometric discord is always lower bounded by the squared Negativity i.e., 
\begin{equation}\label{conjecture} \D\ge\N^2\end{equation} They have rigorously proved the conjecture for arbitrary two-qubit states, all pure states, Werner and isotropic states; besides giving numerical evidences for its validity for all $2\otimes 3$ states. If the conjecture were true, then it would have been perhaps the most interesting operational interpretation via the widely used measure of mixed state entanglement, i.e., negativity, which itself has several direct physical interpretations \cite{VidalWernerPRA02}.  This promising feature has recently been claimed to be analytically true in \cite{Debarbaetalar12}. In this Rapid Communication we show that unfortunately this is not the case.

In proving the conjecture for two-qubit states, the result by Sanpera \etal \cite{SanperaetalPRA98} that the partial transposition (PT) of any two-qubit state can have at most one negative eigenvalue, has played a vital role. It is quite surprising that this decade old interesting result has not been extended to $2\otimes n$ states \cite{AlietalAR07}.  In this work, we extend this result by showing that PT of any $2\otimes n$ state can have at most $(n-1)$ number of negative eigenvalues. While this result is important in its own right, we shall apply it in a systematic way to negate the conjecture and show that there are counterexamples in $2\otimes n$ systems for any $n> 2$. We would like to mention that in a sequel \cite{TufarellietalAr12}, the conjecture has been shown to be  invalid for infinite (hybrid) dimensional systems. However, the present discussion pertains to finite dimensions, for which the conjecture was originally proposed. It is worth pointing out that neither \cite{TufarellietalAr12} nor \cite{PianiAR12} gives any concrete indication  about the purview of the present work, particularly for systems having smaller dimension. For example, it is impossible to deduce that the conjecture will hold for all $2\otimes 2$ states, but not for all $2\otimes 3$ states. We would also like to emphasize that finding a counterexample in $2\otimes 3$ is very difficult. In a generic search scheme (which will be described later in details) we found at most one counterexample among $10^5$ randomly generated states.  Nonetheless, we shall give an explicit example.

We use the same definition of GD and negativity as in \cite{GirolamiAdessoPRA11}, for easy comparison. 
For an $m\otimes n$ $(m\le n)$ state $\rho$, the GD is defined by  \begin{equation}\label{defgd}
\D(\rho)=\frac{m}{m-1}\min_{\chi\in\Omega_0}\|\rho-\chi\|^2=\frac{m}{m-1}\min_{\Pi^A}\|\rho-\Pi^A(\rho)\|^2
\end{equation}
where $\Omega_0$ is the set of zero-discord states (i.e., \emph{classical-quantum states}, given by $\sum p_k|\psi_k\ran\lan \psi_k|\otimes\rho_k$), $\|A\|^2=$ Tr$(A^\dagger A)$ is the Frobenius or Hilbert-Schmidt norm and the last equality is due to \cite{LuoFuPRA10}, the minimization being over all possible von Neumann measurements $\Pi^A=\{\Pi_k^A\}$ on $\rho^A$ and $\Pi^A(\rho):=\sum_k (\Pi_k^A\otimes I^B)\rho(\Pi_k^A\otimes I^B)$. Let $\rho$ have the following Bloch form
\bq\label{blochrho}\rho=\frac{1}{mn}\left[I_m\otimes I_n+\mathbf{x}^t\mathbf{\lambda}\otimes I_n+I_m\otimes\mathbf{y}^t\mathbf{\lambda}+\sum T_{ij}\lambda_i\otimes \lambda_j\right]\eq where $\mathbf{\lambda}=(\lambda_1,\lambda_2,\ldots,\lambda_{d^2-1})^t$ and $\lambda_i$ are the generators of $SU(d)$ for dimension $d=m$ or $n$.  Then a tight lower bound on  GD is given by \cite{RanaParasharPRA12}
\bq\label{lowerboundgd} \D(\rho)\ge\frac{2}{m(m-1)n}\left[\mathbf\|{x}\|^2+\frac{2}{n}\|T\|^2-\sum_{k=1}^{m-1}\lambda_k^{\downarrow}(G)\right]\eq where $\lambda_{k}^{\downarrow}(G)$ are the eigenvalues of $G:=xx^t+\frac{2}{n}TT^t$ sorted in non-increasing order. For all $2\otimes n$ states, equality holds in Eq. \eqref{lowerboundgd} thereby providing an exact (analytic) formula for GD which shall be used later in our analysis.

The Negativity of $\rho$ is given by  \cite{VidalWernerPRA02}
\begin{equation}\label{defneg}
\N(\rho)=\frac{\|\rho^{T_A}\|_1-1}{m-1}=\frac{2}{m-1}\sum_{\lambda_i<0}|\lambda_i(\rho^{T_A})|
\end{equation}
where $\rho^{T_A}$ denotes partial transposition of $\rho$ w. r. t. the system $A$ and $\|X\|_1$ denotes the trace norm given by $\|X\|_1=\tr |X|=\sum|\lambda_i(X)|$. The negativity, as well as its square, are both convex and monotone under local operations and classical communications (LOCC) and hence are legitimate entanglement measure.

\section{\label{any2n} Partial Transposition of any  $2\otimes n$ state has at most $(n-1)$ number of negative eigenvalues}
We begin with the following 
\begin{theorem}
\label{th1}
Partial transposition of any $2\otimes n$ state can not have more than $(n-1)$ number of negative eigenvalues.
\end{theorem} 

\textit{Proof}: This proof is an extension of the $2\otimes 2$ case of Ref. \cite{SanperaetalPRA98}. To proceed with, we want to show (analogous to Theorem $2$ therein) that any hyperplane generated by $n$ orthogonal vectors must contain at least one product vector. Since such a hyperplane is essentially a subspace of dimension $n$, the existence follows from lemma $10$ of Ref. \cite{KrausetalPRA00}. Now, if possible, let the partially transposed state $\rho^{T_A}$ has $n$ (in case of more than $n$, there will be infinite number of product vectors whereas we need the existence of just one product vector to run this proof) number of negative eigenvalues $\lambda_i$ with corresponding eigenvectors $|\psi_i\ran$.  Then expanding the product vector as $|e,f\ran=\sum_{i=1}^nc_i|\psi_i\ran$, we get $$\lan e,f|\rho^{T_A}|e,f\ran=\sum_{i=1}^n\lambda_i|c_i|^2<0$$ But this would imply $\lan e^\ast, f|\rho|e^\ast,f\ran<0$ which is impossible as $\rho$ has to be positive semi-definite.\qed

Note that by Schmidt decomposition any $2\otimes n$ pure state  can be viewed as a pure state in $2\otimes 2$ and hence its partial transposition can have at most one negative eigenvalue.

\section{\label{all23satisfy} Not all $2\otimes 3$ states satisfy $\D\ge \N^2$ }
We will now apply the result of the previous section to prove the following 
\begin{theorem}
\label{th2} There are $2\otimes 3$ states violating  $\D\ge\N^2$.
\end{theorem}

The optimization in our case is quite involved, but nonetheless, we will follow the method  adopted for $2\otimes 2$ case in  \cite{GirolamiAdessoPRA11} (of course, with some modifications) to achieve our goal.  

\textit{Proof:} As noted in \cite{GirolamiAdessoPRA11}, the optimal classical-quantum state $\bar{\chi}$ satisfies $\tr[\bar{\chi}^2]=\tr[\rho \bar{\chi}]$ and hence 
\bq\label{Disexpr}\D=2\|\rho-\bar{\chi}\|^2=2\tr[\rho^2-\bar{\chi}^2]=2\tr[(\rho^{T_A})^2-\bar{\chi}^2]\eq where in the last equality we have used the invariance of Hilbert-Schmidt norm under partial transposition. Let

\begin{subequations}
\label{lambdaxi}\begin{align}
&\lambda_1\ge\lambda_2\ge\lambda_3\ge\lambda_4\ge 0\ge\lambda_5\ge\lambda_6\\
\text{ and }\quad&\xi_1\ge\xi_2\ge\ldots\xi_6\ge 0&
\end{align}
\end{subequations}

be the eigenvalues of $\rho^{T_A}$ and $\bar{\chi}$ respectively. 
Then the Hoffman-Wielandt theorem \cite{HornandJohnsonCUP90} gives  \bq\label{Hoff} \|\rho^{T_A}-\bar{\chi}\|^2\ge\sum_{i=1}^6(\lambda_i-\xi_i)^2\eq
and hence from (\ref{Disexpr}) and (\ref{Hoff}) we have \cite{notequality}, \bq\label{xi2}\sum_{i=1}^6\xi_i^2\le\sum_{i=1}^6\lambda_i\xi_i\eq Besides Eq. (\ref{xi2}), we have the following constraints for a given (fixed) negativity $\N$,
\begin{subequations}\label{othercons}\begin{align}
|\lambda_5|+|\lambda_6|&=\frac{\N}{2}\label{a}\\
\sum_{i=1}^4\lambda_i&=1 +\frac{\N}{2}\label{b}\\
\sum_{i=1}^6\xi_i&=1\label{c}
\end{align}
\end{subequations}

We would now like to explore the following function 
\begin{equation}\label{deff} f(\lambda,\xi):=\sum_{i=1}^6(\lambda_i^2-\xi_i^2)-\frac{\N^2}{2}=\frac{\D-\N^2}{2}
\end{equation} If this function when subjected to the constraints \eqref{xi2} and \eqref{othercons} is always  non-negative, it would indicate support in favour of the conjecture, otherwise it may indicate the existence of a counter example.  

Substituting the value of $\sum \xi_i^2$ from \eqref{xi2} into \eqref{deff}, we have
\begin{equation}\label{modiff} f\ge\sum_{i=1}^6\left(\lambda_i^2-\lambda_i\xi_i\right)-\frac{\N^2}{2}\end{equation}

It can be easily seen (e.g., using Lagrange's multiplier method) that the minimum of $g(x_1,x_2,\dotsc,x_n,y_1,y_2,\dotsc,y_n)$ $=\sum_{i=1}^n(x_i^2-x_iy_i)$ subject to the constraint $x_i\ge 0,y_i\ge 0,\sum_{i=1}^n x_i=a$ and $\sum_{i=1}^n y_i=b$ occurs at $x_i=a/n,y_i=b/n$. So, incorporating  in \eqref{modiff}, the constraints \eqref{b} and \eqref{c} followed by \eqref{a}, we get
\bqas\label{modiff2}f&\ge&\frac{2+\N}{2}\left[\frac{\N}{8}+\frac{\xi_5+\xi_6}{4}\right]+\lambda_5^2+\lambda_6^2+|\lambda_5|\xi_5+|\lambda_6|\xi_6-\frac{\N^2}{2}\\&\ge& \frac{2+\N}{2}\left[\frac{\N}{8}+\frac{\xi_5+\xi_6}{4}\right]+2\left(\frac{\N}{4}\right)^2+\frac{\N}{4}(\xi_5+\xi_6)-\frac{\N^2}{2}\\
&\ge&\frac{\N}{16}\left(2-5\N\right)\eqas  

Thus $f$ is not always non negative. It shows that the conjecture may be violated whenever $\N>2/5$. We note that it does not guarantee the existence of such states, because, the constraints used were just a subset of necessary conditions. This is due the fact that $\bar{\chi}$ is not guaranteed to be optimal classical-quantum state just by these eigenvalue restrictions. Indeed (that is why)  there are several states (the maximally entangled pure state being the easiest example) with more negativity, but satisfying the conjecture. Moreover, in most of the cases, the inequalities are strict and hence nothing can be said conclusively. But the most important consequence of this analysis is that it gives a hint about the existence of some counterexamples. Since it is very difficult to construct a consistent $(\rho,\bar{\chi})$ pair just from an inequality involving their eigenvalues, we are necessarily resorting to numerical techniques. We use the \texttt{QI} package \cite{QIpackage} for this purpose and we found that indeed there are $2\otimes 3$ states violating the conjecture. The code used is given in the supplementary material. We must emphasize that finding the counterexample in $2\otimes 3$ is very difficult. In a typical search, we found at most one counterexample among $10^5$ states (due to limitations on computational power of our desktop PC, we did not try with more states) and we needed to run this code three to four times. In Fig-\ref{ffig1} we show such a successful simulation. For convenience, we have included in the supplementary material an explicit example of a $2\otimes 3$ state violating the conjecture. 
\begin{figure}[ht]
\includegraphics[height=5.3cm]{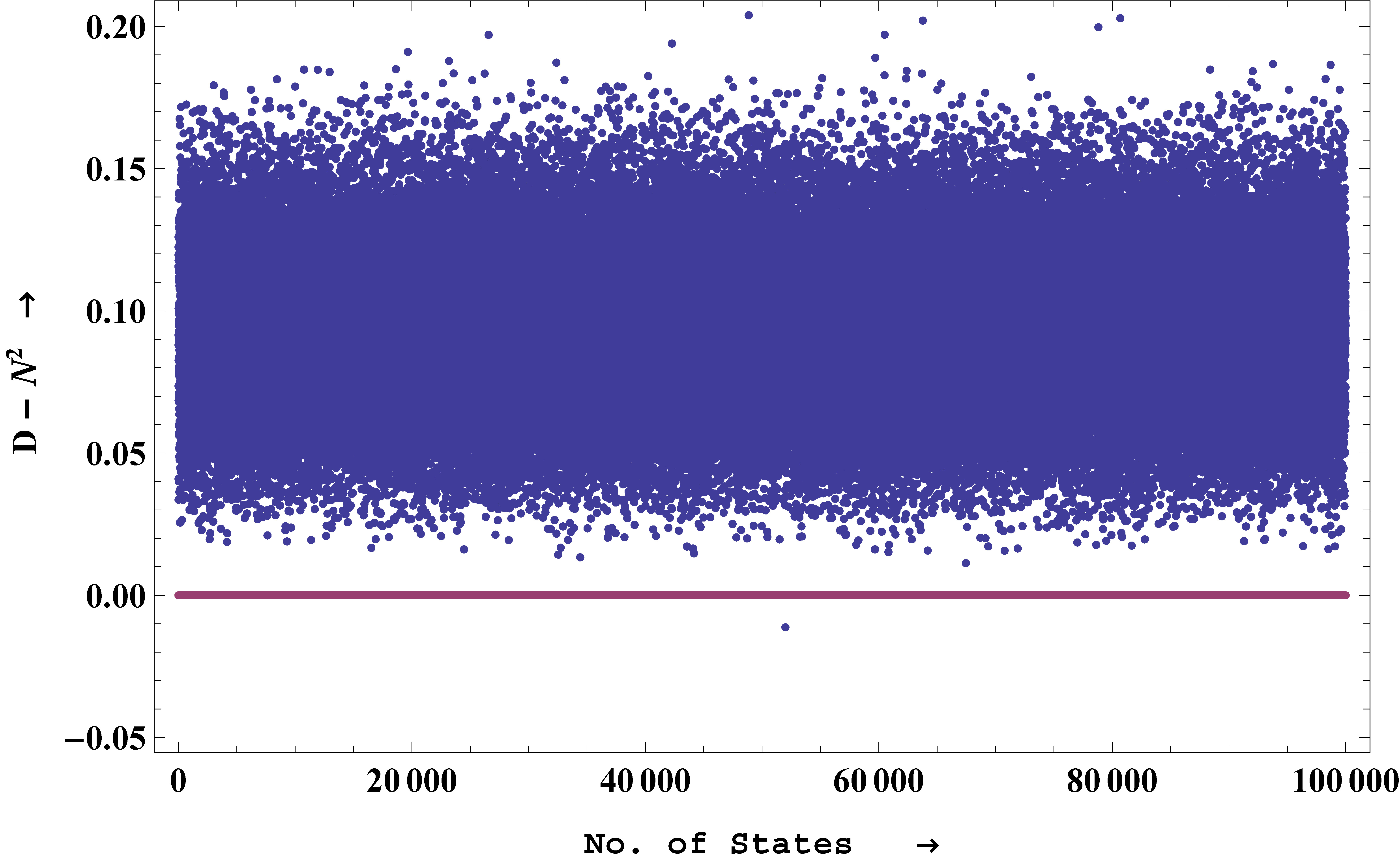}
\caption{(Color online) A successful simulation of the  quantity $\D-\N^2$ (dimensionless) for $10^5$ randomly generated $2\otimes 3$ states using the code given in the supplementary material. The red solid line represents the line $\D=\N^2$. Note that in this simulation we found only one counterexample to the conjecture. }\label{ffig1}
\end{figure}

\section{\label{24count} Counterexamples in $2\otimes n$ dimension for any $n>2$}
\begin{figure}[ht]
\includegraphics[height=5.3cm]{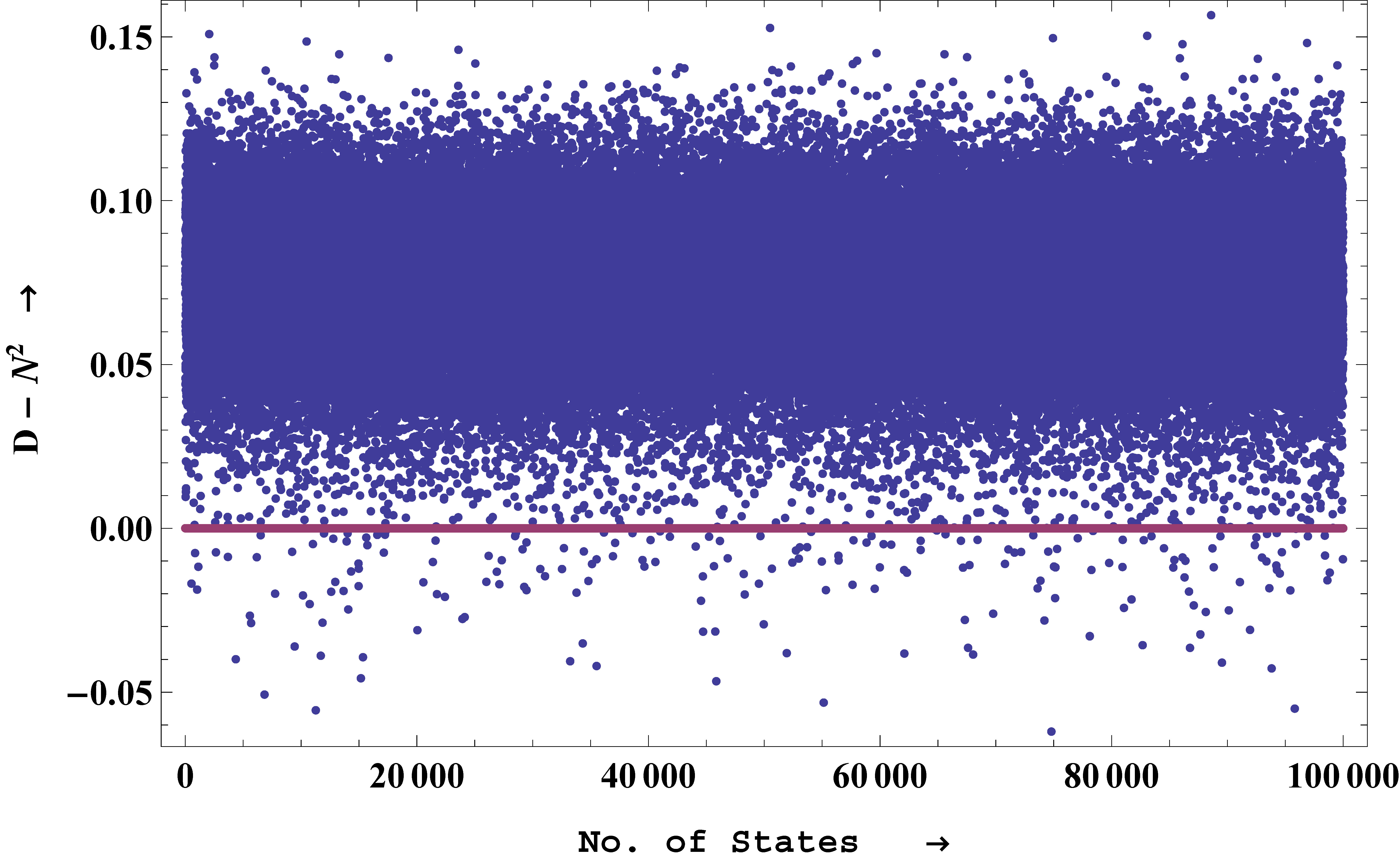}
\caption{(Color online) A simulation of the  quantity $\D-\N^2$ (dimensionless) for $10^5$ randomly generated $2\otimes 4$ states. The red solid line represents the line $\D=\N^2$. The number of states violating the conjecture increases very fast with $n$.}\label{ffig2}
\end{figure}

It is very easy to see that the arguments from previous section also apply to $2\otimes n$ states with $n\ge 3$. It also indicates that the violation ($\D-\N^2$) increases with $n$. Indeed, the number of states violating this conjecture increases very rapidly with $n$. In Fig.-\ref{ffig2} we show a typical simulation of the quantity $\D-\N^2$ for $10^5$ number of randomly generated states in $2\otimes 4$. Using the same method we have seen that in a typical simulation of $10$ randomly generated $2\otimes 15$ states, all states violate this inequality. 

\section{\label{Werner}Counterexample: Werner state in $2\otimes 8$ dimension}
The $m\otimes m$ Werner state is given by \bq\label{wernerstate} \rho_w=\frac{m-z}{m^3-m}\mathbf{I}+\frac{mz-1}{m^3-m}F,\quad z\in[-1,1]\eq
where $F=\sum|k\ran\lan l|\otimes|l\ran\lan k|$. It is well known \cite{LuoFuPRA10} that \bq\label{wernergd44}\D({\rho}_w)= \left(\frac{mz-1}{m^2-1}\right) ^2\eq It has been shown in \cite{GirolamiAdessoPRA11} that all such states satisfy the conjecture \eqref{conjecture}.

For our purpose, we will assume $m=4$. Recently we have shown in \cite{RanaParasharar12} that if we consider the matrix representation of $\rho_w$ (in computational bases) as the density matrix of a $2\otimes 8$ system, the value of $\D$ remains the same. Note that for the calculation of GD, the only relevant quantities are given by $x=0$ and $TT^t=16(1-4z)^2I/225$. Hence $\D_{2\otimes 8}=\D_{4\otimes 4}$ is given by Eq. \eqref{wernergd44} (for completeness, the full derivation has also been presented in the supplementary material).

 Let us now evaluate negativity for the $2\otimes8$ case. With some algebra, the negativity is calculated as 
\bq\label{wernerneg28} \N_{2\otimes8}=\left\{ \begin{array}{ll}
\frac{1}{10} (-2-7 z), & \text{ if } z\in[-1,-\frac{2}{7}) \\
0, & \text{ if } z\in[-\frac{2}{7},\frac{2}{3}] \\
\frac{1}{10} (-2+3 z), & \text{ if } z\in(\frac{2}{3},1]
\end{array}
\right.\eq

\begin{figure}
 \includegraphics[height=5.3cm]{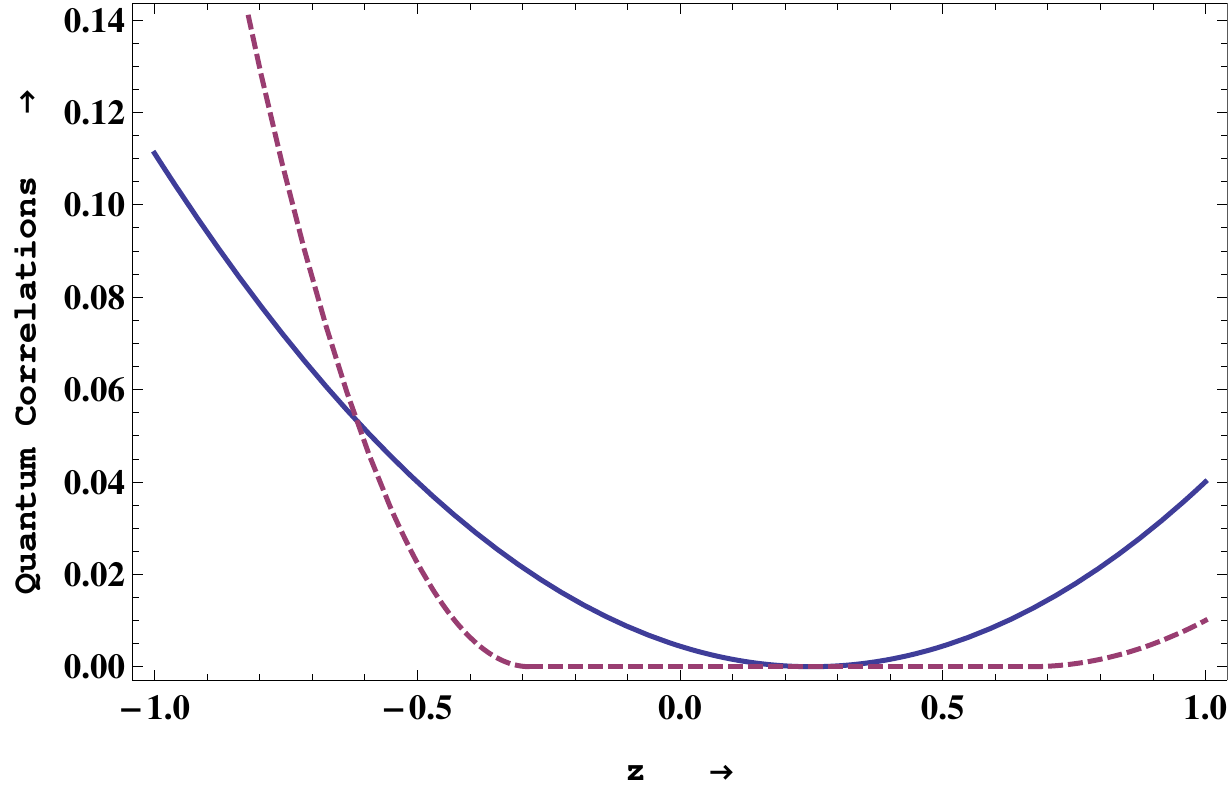}
 \caption{(Color online) Geometric Discord ($\D$, shown by solid blue line) and Squared Negativity ($\N^2$, shown by dashed red line) for the $2\otimes 8$ dimensional Werner states. We note that for all $z\in[-1,-8/13)$, the state violates the conjecture. All quantities plotted are dimensionless.}\label{ffig3}
 \end{figure}

 Thus \bqs \D_{2\otimes8}<\N_{2\otimes8}^2\eqs holds for all $z\in[-1,-8/13)$. In all other cases, Eq. (\ref{conjecture}) is satisfied. The GD and  Squared Negativity for this $2\otimes 8$ case is shown in Fig-\ref{ffig3}.
 
We note that though the isotropic state also satisfies $\D_{2\otimes8}=\D_{4\otimes4}$ \cite{RanaParasharar12}, it does not violate Eq. (\ref{conjecture}).

\section{\label{conclusion}Discussion and conclusion} 
In this work we have negated a recently made conjecture about the interplay between geometric discord and negativity. In the process, we have used the fact that partial transposition of any $2\otimes n$ state can not have more than $(n-1)$ number of negative eigenvalues; an extension of the well known two-qubits result. Two possible issues are yet to be studied: i) To our knowledge, an upper bound on the number of negative eigenvalues of partial transposition of generic $m\otimes n$ state is not known, ii) We have given just an upper bound for $2\otimes n$ state. Our numerical exploration suggests that with increasing $n$, the number of states satisfying this bound  decreases very fast. We have seen that this bound is saturated for $n\le 4$. The situation for higher dimensional systems is not clear.

Presently, we have restricted our attention to the $2\otimes n$ case as the analytic formula for GD is known only for such states. Though the main purpose of this work has been solved with $2\otimes n$ states, we do not know about $m\otimes m$ systems (with finite $m$) but hopefully there will be some counterexamples. Also it does not look too difficult to find such examples using the generic bound developed in \cite{RanaParasharPRA12}. Thus, we have shown that beyond the very special case of two-qubits, there is no such linear interrelation between geometric discord and squared negativity. Probably the same holds for other measures of correlations as well. As has been pointed out in \cite{PianiAR12}, the geometric discord, as a measure of quantum correlations, should be used with care. Finally, though our result tends to support the objection raised against the defining notion of geometric discord, we do believe the measure itself is indeed useful, due to its several novel  operational interpretations in terms of optimal resource in various physical tasks.


\begin{thebibliography}{40}
\bibitem{notequality} We note that this should be an inequality constraint ($\le$) rather than equality as in \cite{GirolamiAdessoPRA11}. Though equality holds for pure states and Werner states (possibly with other classes of states), even for generic $2\otimes 2$ states, equality does not hold. Our $2\otimes 3$ counterexample also satisfies strict inequality. It is worth mentioning that the minimum point corresponding to $\lambda_1=\lambda_2=\lambda_3,\xi_1=\xi_2=\xi_3=1/3, \xi_4=0$ in  \cite{GirolamiAdessoPRA11} does not satisfy this equality but rather the inequality.
\end{thebibliography}
\end{document}